\documentclass[twocolumn,showpacs]{revtex4}
\usepackage{epsfig}
\newcommand{\bm}[1]{ \mbox{\boldmath $#1$}  }

\begin{document}

\title{ Isomeric $0^{-}$ halo-states in $^{12}$Be and $^{11}$Li}

\author{C. Romero-Redondo}
\author{E. Garrido} 
\affiliation{ Instituto de Estructura de la Materia, CSIC, 
Serrano 123, E-28006 Madrid, Spain }
\author{D.V. Fedorov}
\author{A.S.~Jensen}
\affiliation{ Department of Physics and Astronomy,
        University of Aarhus, DK-8000 Aarhus C, Denmark }

\date{\today}

\begin{abstract}
We predict the existence of an isomeric 0$^-$-state in $^{12}$Be at an
excitation energy of about $2.5$~MeV, and a 0$^-$-resonance in
$^{11}$Li with both energy and width of about $1$~MeV corresponding to
two-neutron emission.  The structure of these halo-like states are
like the 1$^-$-states which means essentially a core surrounded by two
neutrons in single-particle $s$ and $p$-states. The life-time of the
$^{12}$Be state is determined by $M1$ or $M2$-emission, $\tau(M1)
\approx 10^{-11}$~s or $\tau(M2) \approx 10^{-8}$~s estimated for photon energies 
of $0.1$~MeV and $0.6$~MeV, respectively. 
\end{abstract}


\maketitle

\paragraph*{Motivation.}
Although light nuclei exhibit amazingly individual characters
\cite{suz03}, identical structures can show up in these otherwise very
different quantum systems. A prominent example are the isobaric analog
states which in principle can be traced through a sequence of
neighboring nuclei \cite{boh69}. Another type of similarity appears
for cluster states when the corresponding threshold is approached
\cite{ike68}.  This is seen for halo nuclei of two and three-body
character with the Borromean two-neutron nuclei $^{6}$He and $^{11}$Li
as well-known examples \cite{jen04}.

The individual characters of light nuclei are reflected in relatively
large changes of structure by addition of one or a few nucleons.
Substantial efforts have been devoted to investigate the changing
shell structure as the driplines are approached \cite{iwa00,pai06}.
In particular the $N=8$ neutron shell is very stable at the
$\beta$-stability line for $^{16}$O while it has disappeared for
$^{12}$Be and $^{11}$Li.

This was recognized very early by Talmi and Unna \cite{tal60} who
traced the $1p_{1/2}$ and $2s_{1/2}$ neutron levels from $^{13}$C to
$^{11}$Be.  These levels, belonging to different harmonic oscillator
shells, approach each other and eventually cross when the neutron
excess increases.  Close to the neutron dripline they both appear just
above the Fermi energy.  This phenomenon, known as parity inversion,
explains the fact that a nucleus like $^{13}$C has a $1/2^-$ ground
state, while the ground state of the one-neutron halo nucleus
$^{11}$Be is 1/2$^+$ \cite{ajz90}. For 9 neutrons different ground state spins,
5/2$^+$ and 1/2$^+$, are also found for $^{17}$O and $^{15}$C. For the particle 
unstable nucleus
$^{10}$Li different theoretical works predicted the existence of a
similar low-lying $s$-wave intruder state \cite{bar77,joh90,tho94}, in
such a way that the ground state in $^{10}$Li should correspond to a
state with negative parity (the ground state of $^9$Li has spin and
parity 3/2$^-$). The available experimental data concerning the ground
state properties of $^{10}$Li are however controversial, although most
of them point towards the existence of such a low-lying virtual
$s$-state \cite{kry93,you94,abr95,zin95}. This result has also been
recently supported in \cite{jep06}.

Having in mind the structure of the nuclei in the $N=7$ isotonic
chain, one could wonder about the properties of the spectrum for the
nuclei in the $N=8$ chain. Independently of the existence of an
intruder $s$-state, it is clear that some of the excited states
of these nuclei should arise from the excitation of the neutrons
inside the $sp$-shell. In particular, when one of the neutrons is in
the $s_{1/2}$-shell and the other one in the $p_{1/2}$-shell, this
structure can obviously lead to either a 1$^-$ or a $0^-$ excited
state. This doublet can actually be found for instance in $^{14}$C
with excitation energies of 6.09 MeV and 6.90 MeV, respectively
\cite{ajz91}, and in $^{16}$O with excitation energies of 9.59 MeV and
10.97 MeV, respectively \cite{ajz86}. A linear extrapolation in mass 
from $^{16}$O over $^{14}$C to $^{12}$Be 
and $^{11}$Li leads to predictions of (2.8 MeV,2.6 MeV) and (0.80 MeV,0.84 MeV)
for $^{12}$Be and $^{11}$Li, respectively for $0^-$ and $1^-$. For $^{16}$N and 
$^{16}$F, for which some of the states should correspond to one neutron and 
one proton in the $sp$-levels, the spectra also present the same $0^-$, $1^-$ states 
at (0.12 MeV,0.40 MeV) and (ground state, 0.19 MeV). The 0$^-$ is close but 
below the $1^-$ state, as repeated in $^{14}$N but with larger spacing.

For all these reasons, it is surprising that for $^{12}$Be and
$^{11}$Li, also belonging to the $N=8$ chain, information is
available only about the 1$^-$ excited states. For $^{12}$Be a bound
$1^-$ state has been found with excitation energy of 2.68 MeV
\cite{iwa00}, while for $^{11}$Li experimental evidences about the 
existence of an unbound $1^-$ excited state has also been given
\cite{kor97}. No predictions have been formulated about the occurrence
of 0$^-$-states. Only in \cite{kan03} an unbound $0^-$ state was
predicted for $^{12}$Be. However in that
work the energies of the negative parity states are systematically
overestimated, i.e. the $1^-$ state is also clearly unbound, contrary
to the experimental knowledge. Also in shell model calculations the
simultaneous inclusion of different parities is a general source of
uncertainty due to the requirement of a larger Hilbert space. This problem
is still present in recent no-core shell model results obtained for $^{11}$Be
\cite{for05}.

In general high-lying $0^{-}$ and $1^{-}$-states of short half-life
are abundant throughout the chart of nuclei.  On the other hand, such
low-lying states are rare especially if they are of simple
single-particle structure and with a long lifetime classifying them as
isomeric states.  The purpose of this letter is to present
theoretical evidence for the existence of so far unknown (isomeric)
0$^-$-states in both $^{12}$Be and $^{11}$Li.  We shall estimate some
of the properties of these states and especially predict energies and
transition strengths.  It turns out that the most probable energy for
$^{12}$Be($0^{-}$) is below the particle emission threshold with
magnetic dipole or quadrupole transitions as the only possible decay
channels.  Its lifetime should therefore be comparable to the recently
discovered isomeric $0^{+}$-state \cite{shi03}.

\paragraph*{Theoretical formulation.}

We use a three-body model to describe both nuclei, with a $^{10}$Be or
$^{9}$Li core surrounded by two neutrons. In \cite{nun02} the role
played by core excitations in $^{12}$Be is investigated.  The lowest
excited state in $^{10}$Be is a 2$^+$ state, which can not couple
the two neutrons in the $s_{1/2}$ and $p_{1/2}$-states to total
angular momentum zero. Therefore, the $^{10}$Be excitations should not
contribute significantly in this case and we can then assume an inert
core. The three-body wave functions are obtained by solving the
Faddeev equations with the hyperspheric adiabatic expansion method
\cite{nie01}. Unbound resonant states are computed by using the 
complex scaling method \cite{fed03}.

For the $^{10}$Be-neutron interaction we have constructed a simple
$\ell$-dependent gaussian potential containing central and spin-orbit
terms. The range of the gaussians has been chosen equal to 3.5 fm.
For $s$-waves a strength of $-8.40$ MeV places then the 1/2$^+$-state 
in $^{11}$Be at $-0.504$ MeV, that matches the experimental value
\cite{ajz90}. The use of this shallow $s$-wave potential is an
efficient way of taking into account the Pauli principle, since the lowest
$s_{1/2}$-shell is fully occupied in the $^{10}$Be-core. This procedure is
phase-equivalent to use a deeper potential, binding the neutrons in the 
Pauli forbidden $s$-state, and afterwards removing it from the active space 
available for the three-body system \cite{gar97}. For $p$-waves the strengths of the central and
spin-orbit gaussians are 40.0 MeV and 63.52 MeV, which produces a
bound $1/2^-$-state in $^{11}$Be at $-0.184$ MeV, in agreement with
the experiment \cite{ajz90}.  The $p_{3/2}$-wave is at the same time
pushed up to high energies, where it remains unoccupied as it should
since this state already is occupied by core-neutrons and therefore
Pauli forbidden. This is achieved by using an inverted $p$-wave spin-orbit force.

For the $^{12}$Be calculation we have also included a $d$-wave
potential, giving rise to $5/2^+$ and $3/2^+$-resonances at 1.28 MeV
and 2.90 MeV, respectively, above threshold, which again match 
the experimental values \cite{ajz90}.  This is obtained by using
gaussian $d_{5/2}$ and $d_{3/2}$-potentials with strengths equal to
$-43.8$ MeV and $-199$ MeV, respectively. For the $d_{3/2}$ gaussian
potential a range of 1.7 fm (instead of 3.5 fm) has been used in order
to produce a narrower resonance in better agreement with the
experiment.  The neutron-neutron interaction is from \cite{gar99}.

For the $^{11}$Li calculations we use the simple $^9$Li-neutron
interaction quoted as potential IV in table I of \cite{gar99}.  A more
sophisticated potential, like the one used in \cite{gar97}, where the
Pauli principle is accounted for by use of phase equivalent potentials
could be used, but as shown in \cite{gar99b} the results are
indistinguishable. Furthermore potential IV in table I of \cite{gar99}
can be used assuming zero spin for the $^9$Li-core, which permits us
to perform a calculation fully analogous to the one for $^{12}$Be, and
at the same time we can investigate the $0^-$-state directly without
the entanglement of the coupling to the core-spin of $3/2$. This is as 
realistic as the full complication of including the $^9$Li
core-spin of 3/2 \cite{gar99}, where the coupling of the 3/2$^-$ core-state to 
a two-neutron 0$^-$-state leads to a 3/2$^+$ state. For
$d$-waves we use the same potential as for $p$-waves.  The
neutron-neutron interaction is again from \cite{gar99}.

\begin{table}[tbh]
\begin{scriptsize}
\caption{Components included in the calculations for the 1$^-$-states. 
The left and right parts correspond to the first Jacobi set ($\bm{x}$
between the two neutrons) and the second and third Jacobi sets
($\bm{x}$ from core to neutron), respectively. The last row gives the
maximum value of the hypermomentum used for each component. The
components written in bold letters are used in the
calculation of the $0^-$ states.\\}
\begin{tabular}{|c|ccccc|cccccccccc|}
\hline
 $\ell_x$ & {\bf 1} &  0 &  2 & {\bf 1} &  1 &  1 & {\bf 1}  &  0 & {\bf 0}  &  1 & {\bf 1}  &  1 &  2 & {\bf 2}  &  2  \\
 $\ell_y$ & {\bf 0} &  1 &  1 & {\bf 2} &  2 &  0 & {\bf 0}  &  1 & {\bf 1}  &  2 & {\bf 2}  &  2 &  1 & {\bf 1}  &  1  \\
 $L$      & {\bf 1} &  1 &  1 & {\bf 1} &  2 &  1 & {\bf 1}  &  1 & {\bf 1}  &  1 & {\bf 1}  &  2 &  1 & {\bf 1}  &  2  \\
 $s_x$    & {\bf 1} &  0 &  0 & {\bf 1} &  1 & 1/2& {\bf 1/2}& 1/2& {\bf 1/2}& 1/2& {\bf 1/2}& 1/2& 1/2& {\bf 1/2}& 1/2 \\
 $S$      & {\bf 1} &  0 &  0 & {\bf 1} &  1 &  0 & {\bf 1}  &  0 & {\bf 1}  &  0 & {\bf 1}  &  1 &  0 & {\bf 1}  &  1  \\
 $K_{max}$& 119     & 99 & 61 & 81      & 61 & 99 & 119      & 99 & 119      & 41  & 41      & 41 & 41 & 41       & 41  \\
\hline
\end{tabular}
\label{tab1}
\end{scriptsize}
\end{table}

\paragraph*{Results.}

When the interactions above are used, the $0^+$ ground states of
$^{12}$Be and $^{11}$Li are obtained with two-neutron separation
energies equal to $-3.67$ MeV and $-0.30$ MeV, respectively, both of
them matching the experimental values \cite{you93,ajz90}. For $^{11}$Li 
the agreement with the experiment has been obtained after using a gaussian
three-body force with a range of 2.5 fm and
a strength of $-4.0$ MeV.
All components with $\ell_x=\ell_y=0,1,2$ are included. The weights of
the $s$, $p$, and $d$-waves in the core-neutron channel turn out to be
74\%, 15\%, and 11\% for $^{12}$Be, and 64\%, 35\%, and 1\% for
$^{11}$Li, respectively.

For $^{12}$Be particle-bound $0^+$ and 2$^+$ excited states are found
with two-neutron separation energies of $-0.59$ MeV and $-0.62$
MeV. These two states are experimentally known to be present in the
$^{12}$Be spectrum with binding energies $-1.43\pm0.02$ MeV
\cite{shi03} and $-1.56\pm0.01$ MeV \cite{iwa00b}, respectively. 
Therefore an effective three-body force is here needed to recover the
experimental values.

\begin{table}[tbh]
\caption{Energy $E$ of the bound $0^-$ and $1^-$-states in $^{12}$Be 
and energy and width $(E_R,\Gamma_R)$ of the $0^-$ and $1^-$-states in
$^{11}$Li. The last two rows give the percentage of the three-body
wave function corresponding to neutron and core in a relative $s$- or
$p$-wave.\\}
\begin{tabular}{|c|cc|cc|}
\hline
     &   $1^-$  &  $0^-$  & $1^-$  &  $0^-$  \\
\hline
$E$ or $(E_R,\Gamma_R)$ & $-0.97$ & $-0.96$ & $(0.64,0.32)$ & $(0.92,0.82)$ \\
$\%s$-wave & 51\%  & 54\% & 43\% &  60\% \\
$\%p$-wave & 49\%  & 46\% & 57\% &  40\% \\
\hline
\end{tabular}
\label{tab2}
\end{table}

In table \ref{tab1} we give the components included in the
calculations for the 1$^-$-states. For $^{12}$Be a bound 1$^-$-state
is found with a binding energy of $-0.97$ MeV, which agrees with the
experimental value of $-0.99\pm0.03$ MeV \cite{iwa00}. For $^{11}$Li a
$1^-$-resonance is found with energy and width
$(E_R,\Gamma_R)$=$(0.64,0.32)$ MeV, which also agrees with the
experimental excitation energy of about 1 MeV given in
\cite{kor96,gor98} (the ground state in $^{11}$Li is bound with a two
neutron separation energy of 0.3 MeV). These values have been obtained
without inclusion of effective three-body forces.

The components written in bold letters in table~\ref{tab1} are used in
the computation of the $0^-$-states.  For $^{12}$Be a bound
$0^-$-state is found with a two-neutron separation energy of $-0.96$
MeV, which corresponds to an excitation energy roughly 10 keV higher than for
the $1^-$-state.  For $^{11}$Li a $0^-$-resonance has also been found,
with energy and width $(E_R,\Gamma_R)$=$(0.92,0.82)$ MeV, which
corresponds to an excitation energy roughly 300 keV higher than for
the $1^-$-resonance. In table~\ref{tab2} we summarize the results
corresponding to the $0^-$ and $1^-$-states for both nuclei.  The
labels $\%s$ and $\%p$ refer to the weights in the three-body wave
function of the components corresponding to the core and one of the
neutrons in a relative $s$ or $p$-wave. Although $d$-waves are
included in the calculation, their contributions to the $0^-$ and
$1^-$-states are negligible.To test the robustness
of the results for the 0$^-$ states we have repeated the calculations
for $^{12}$Be using the Argonne $v_8$ neutron-neutron potential. The 0$^-$ and
$1^-$ states appear then more bound by almost 200~keV and 150~keV,
respectively.

\begin{figure} [h]
\epsfig{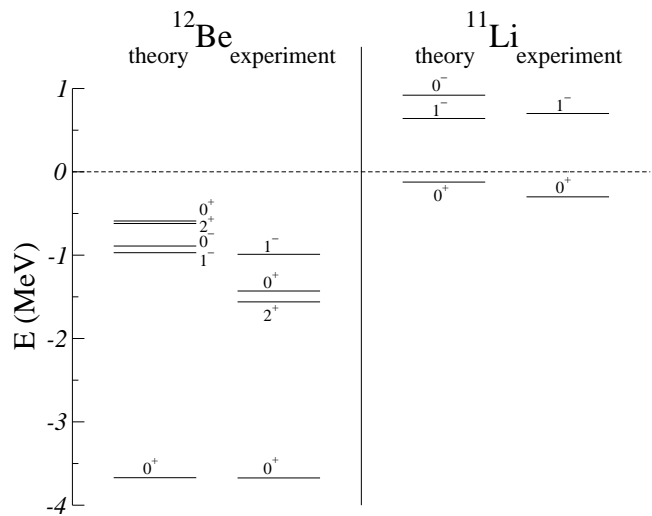}
\caption[]{The computed (without effective three-body forces)
and experimental low-lying spectra of $^{12}$Be 
and $^{11}$Li obtained in two-neutron cluster models with
neutron-core $s$, $p$ and $d$-waves. }
\label{fig1}
\end{figure}

The low-lying spectra are sketched in fig.~\ref{fig1} for both
$^{12}$Be and $^{11}$Li. The results without adjustment by the
three-body interaction are compared to the established experimental
energies. For $^{12}$Be the excited $0^+$ and $2^+$-states both appear
above the experimental results while the $1^-$-state agrees with the
measured value.  The $0^-$-state is built of the same levels as the
$1^-$-state and it therefore is a strong indication that also the
predicted $0^-$-state is particle-bound.  In any case it would be very
unusual to miss these three-body states by 1~MeV as required to make
it unbound. Even in that unlikely event there should be a resonance
structure as predicted in $^{11}$Li.  However, for $^{12}$Be the
spectrum should be much cleaner than for $^{11}$Li where the low-lying
continuum already is crowded by the three different $1^-$-excitations
on top of the $3/2^-$ ground state.

\paragraph*{Wave functions and transition strengths.}

\begin{figure} [h]
\epsfig{file=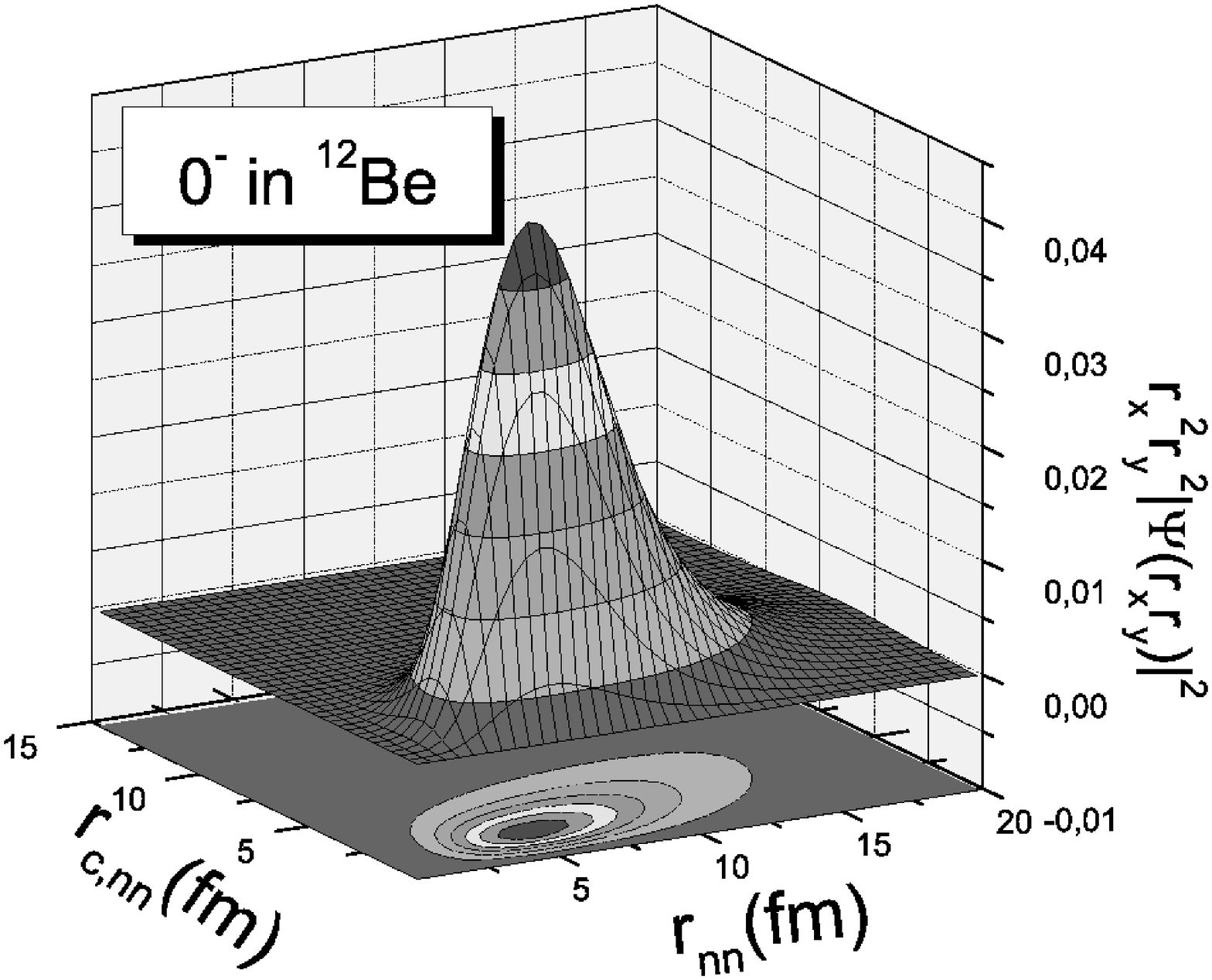, width=6.5cm, angle=0}
\epsfig{file=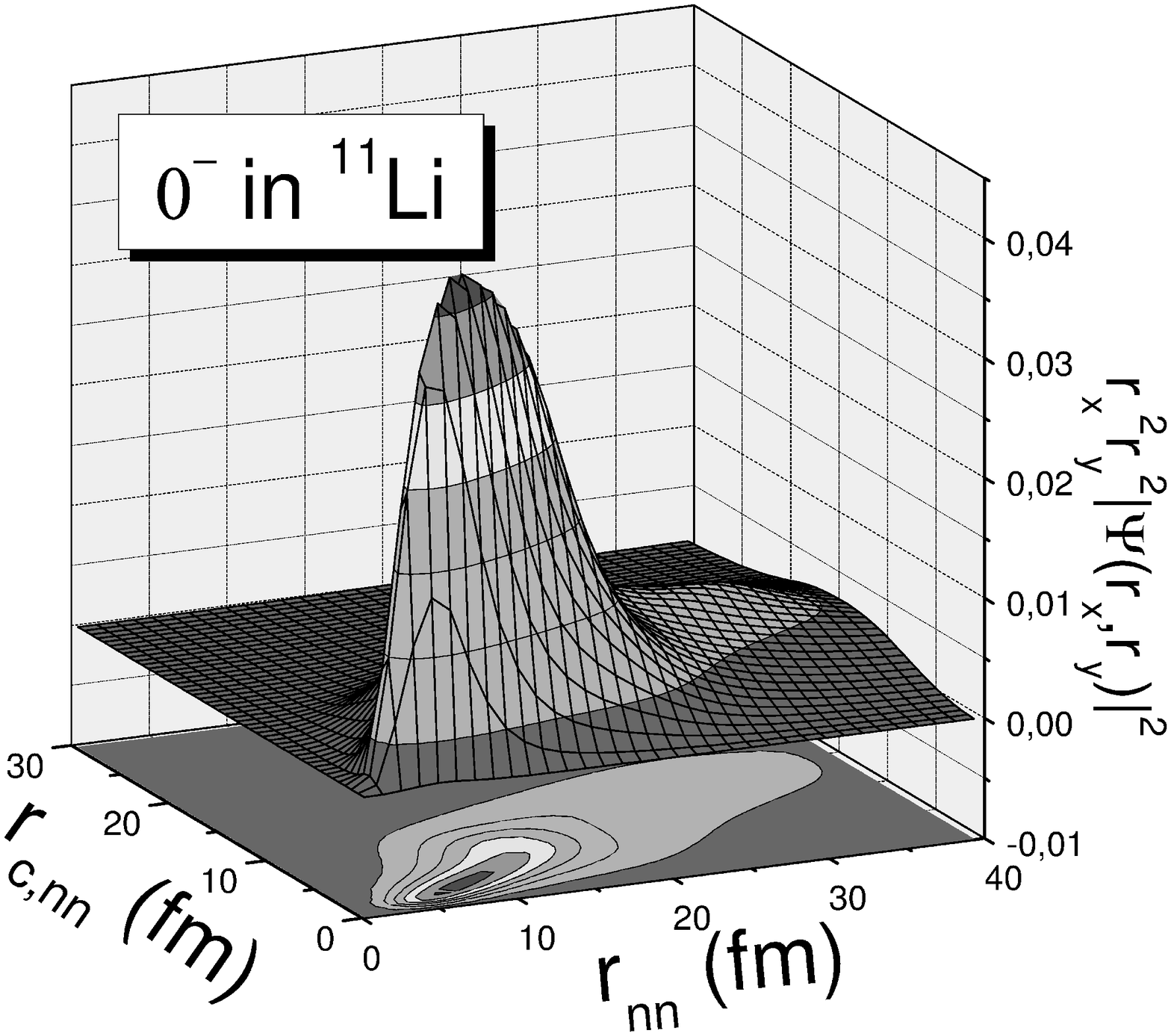, width=6cm, angle=0}
\caption[]{The probability distributions of the $0^-$-states for
$^{12}$Be (upper part) and $^{11}$Li (lower part) as functions of the
distances between the two neutrons and their center-of-mass and the
cores.  The wave functions are computed in two-neutron cluster models
with neutron-core $s$, $p$ and $d$-waves. }
\label{fig2}
\end{figure}

The relatively small binding energy allows halo formation.  The
dimensionless quantity $\langle \rho^2\rangle/\rho_0^2$ ($\rho$ is the
hyperradius) used as criterion in \cite{jen04} is 2.4 and 2.5 for the
$0^-$-states in $^{12}$Be and $^{11}$Li, i.e. both numbers are larger
than 2, indicating halo configuration.  For the $^{11}$Li-resonance
the number quoted is $\left[Real\left(\frac{\langle\rho^2
\rangle^{1/2}}{\rho_0}\right)\right]^2$, where $\langle \rho^2
\rangle$ is complex and $\rho_0$ is real as defined in \cite{jen04}).
The density distributions can be seen in the upper ($^{12}$Be) and
lower ($^{11}$Li) parts of fig.~\ref{fig2}. For $^{11}$Li the complex
scaled three-body resonance wave function is shown ($\theta=0.3$
rads). Both the $0^-$-states in $^{12}$Be and $^{11}$Li have the
largest probabilities when the two neutrons are well separated by
about 5 fm and 8 fm, respectively, and the cores are close to the
two-neutron center-of-mass by about 2 fm. 

For these $0^-$-states, the average root-mean-square distances between
pairs of particles are $\langle r_{nn}^2 \rangle^{1/2}$= 9.2 fm and
$\langle r_{nc}^2 \rangle^{1/2}$= 5.9 fm for the neutron-neutron and
neutron-core distances in $^{12}$Be, and $\langle r_{nn}^2
\rangle^{1/2}$= 8.7 fm and $\langle r_{nc}^2 \rangle^{1/2}$= 5.1 fm
for the same distances in $^{11}$Li.

The possibility of detecting the $0^-$-state in $^{12}$Be might depend
strongly on its life-time which in turn depends on the allowed decay
modes.  Since the relative energies are uncertain we have to compute
decays to both $1^-$ and $2^+$-states which means magnetic dipole or
quadrupole transitions. The operators are
\begin{eqnarray}   \label{e20}
 {\cal M}_\mu(M1) &=& \frac{e\hbar}{2Mc}\sqrt{\frac{3}{4\pi}} 
\sum_i (g^{(i)}_s \vec s_i  + g^{(i)}_\ell \vec \ell_i)_{\mu}  \\
 {\cal M}_\mu(M2) &=& \frac{e\hbar}{Mc} {\frac{5}{\sqrt{2}}} 
\sum_i \sum_{\nu,q} 
\left(
    \begin{array}{ccc}
         1 & 1 & 2 \\
       \nu & q & -\mu
    \end{array}
    \right)
Y_{1,\nu}(\Omega_i)  \nonumber \\
&&
\left( 
g^{(i)}_s \vec s_i  + \frac{2 g^{(i)}_\ell}{3} \vec \ell_i
\right)_q\; ,
\end{eqnarray}
where the constants $g_s$ and $g_\ell$ depend on the constituent
particles $i$. We can identify three different sources of
uncertainties in the lifetime estimates, i.e.  (i) the effective
values of the g-factor in these expressions, (ii) the precise values
of the excitation energies or rather the emitted photon energy, and
(iii) contributions from degrees of freedom beyond the three-body
model.  We believe that uncertainties due to (iii) are relatively
small and consider in the following only effects of three-body
structures.  When the excitation energies are fine-tuned to reproduce
experimental information by use of three-body potentials we
essentially maintain the three-body structures and thereby the matrix
elements.

The $0^-$-state has a very similar composition to the $1^-$-state
which reproduces the experimental energy without any three-body force.
This indicates that the computed $0^-$-energy also is close to the
correct value.  The lifetime estimates are then reliable determined by
the corresponding matrix elements and the relative energies.  Thus the
$0^-$-state can decay to both the $1^-$ and $2^+$-states by magnetic
dipole and quadrupole emission, respectively.

\begin{table}
\caption{The computed magnetic transitions, ${\cal B}(M1)$ (in $e^2 fm^2$) 
and ${\cal B}(M2)$ (in $e^2 fm^4$) for $^{12}$Be. The core values are
$g_\ell^{(c)} = 4$, $g_s^{(c)}=0$. \\ }
\begin{tabular}{|c|c|c|c|}
\hline
$g_\ell^{(n)}$ & $g_s^{(n)}$ &   ${\cal B}(M1,0^-\rightarrow 1^-)$ &  ${\cal
B}(M2,0^-\rightarrow 2^+)$ \\ \hline
$0.00$ & $-2.00$ & 0.014  &  0.235  \\
       & $-3.82$ & 0.052  &  0.858  \\ \hline
$0.15$ & $-2.00$ & 0.016  &  0.257  \\
       & $-3.82$ & 0.056  &  0.899  \\ \hline
$0.28$ & $-2.00$ & 0.018  &  0.277  \\
       & $-3.82$ & 0.059  &  0.936  \\ \hline

\end{tabular}
\label{tab3}
\end{table}

The core has angular momentum zero and therefore a vanishing effective
spin $g_s^{(c)}$-factor and $g_\ell^{(c)} = 4$.  We also use the free
value of $g_s^{(n)} = -3.826$ and we calculate an effective charge 
corresponding to
$g_\ell^{(n)} = 0.28$ which reproduces the $^{11}$Be transition
strength ${\cal B}(E1,1/2^-\rightarrow 1/2^+)=0.115\pm0.010$ $e^2fm^2$
\cite{mil83}.  The transition operators are then defined and we can
compute ${\cal B}(M1,0^- \rightarrow 1^-)$ and ${\cal B}(M2,0^-
\rightarrow 2^+)$.  However, the effective values of these $g$-factors
are rather uncertain and spin polarization could reduce $g_s$ by a
factor of 2, change $g_\ell^{(c)}$ by perhaps 10~\%, and vary
$g_\ell^{(n)}$ from the assumed effective value, see also the
discussion of the empirical evidence in \cite{boh69}.

In practice the present magnetic dipole transition is determined by
the motion of the neutrons because the difference in the two states
$0^-$ and $1^-$ is essentially a spin flip of one of the neutrons.
The contribution from the core is very small, first because it has 
spin zero, and second, because due to its large mass compared to the 
mass of the neutrons, the orbital angular momentum of the core from the 
three-body center of mass is also small.  The
dependence on the neutron $g$-factors are seen in table~\ref{tab3}.
The dependence on the orbital part is a lot smaller than the spin part
arising from $g_s^{(n)}$.  In total with a rather wide interval of
parameter variation the changes are within a factor of 3 and certainly
within an order of magnitude in this three-body model.

The lifetimes of the corresponding two possible decay modes are given
by trivial factors and specific power law dependences on energy
\cite{boh69}.  With the values from table~\ref{tab3} obtained for
$g_\ell^{(n)} = 0.28$, $g_s^{(n)}=-3.82$ we find
\begin{eqnarray}   \label{e60}
   \tau(M1)\approx \bigg( \frac{0.1~{{\rm MeV}}}{\hbar \omega_1}\bigg)^3 
   1.06 \cdot10^{-11} \;\; \mbox{s} \; ,  \\
 \tau(M2)\approx \bigg(\frac{0.6~{{\rm MeV}}}{\hbar \omega_2}\bigg)^5
  1.12\cdot10^{-8} \;\; \mbox{s} \; ,
\end{eqnarray}
where the emitted photon energies are $\hbar \omega_1$ and $\hbar
\omega_2$.  The uncertainties due to the $g$-factors are easily found 
by combining table~\ref{tab3} and eq.(\ref{e60}).  Since $\hbar
\omega_2$ is expected to deviate much less than $\hbar \omega_1$ from 
their correct values we deduce that $\tau(M1)$ decides the lifetime of
the $0^-$-state, unless the energies of the $0^-$ and $1^-$-states
coincides to a remarkable accuracy of less than 10~keV.  Thus the
lifetime is estimated to be longer than about $10^{-11}$~s and shorter than
about $10^{-8}$~s which qualifies the name of an isomeric state.
Due to the uncertainties arising from the three-body forces the $0^-$
state could be below the $1^-$ level, and there is even a small probability
for being unbound. In case of being below the $1^-$, then only the $M2$
transition to the $2^+$ state would be allowed and the lifetime would be
confined to be around $10^{-8}$~s.  If the $0^-$-state is above the
two-neutron threshold, the state should resemble the
corresponding resonance in $^{11}$Li. In that case the width should
be comparable to the resonance energy, e.g. an energy below say 0.2 MeV
above threshold would be related to a width smaller than that value.

\paragraph*{Summary and conclusions.}

We predict a particle-bound isomeric $0^-$-state in $^{12}$Be with a
dominant configuration of two neutrons around the $^{10}$Be-core.  A
similar low-lying but particle-unbound $0^-$-resonance of rather large
two-neutron emission width is predicted in $^{11}$Li. The decay of
$0^-$ in $^{12}$Be has to be by magnetic dipole or quadrupole photon
emission.  The lifetime $\tau$ is estimated with large uncertainty to
be in the interval $10^{-11} - 10^{-8}$~s.  
This 0$^-$ state of $^{12}$Be can be reached from the ground state by a single
particle excitation from an $s_{1/2}$ to a $p_{1/2}$ state. It has a
structure essentially only deviating from the known 1$^-$ state by
neutron spin couplings to zero instead of 1. This $^{12}$Be-state can be 
expected to be populated in reactions by a cross
section about the statistical factor of three smaller than that known for
the similar $1^-$-state.  Due to the long lifetime of the isomeric
$0^-$-state of $^{11}$Li its production rate from the $^{11}$Li ground
state can be expected to be much weaker than for the $1^-$-state.

\paragraph*{Acknowledgment.}

We highly appreciate discussions with Dr. H.O.U. Fynbo in the initial
phase of the present investigation. This work was partly supported by 
funds provided by DGI of MEC (Spain) under contract No. FIS2005-00640.
One of us (C.R.R.) acknowledges support by a predoctoral I3P fellowship
from CSIC and the European Social Fund.

\end{document}